\def\danny{Piper Jaffray, LLC, 245 Park Ave., 33rd floor, New York, New York 10167, USA}
\def\ias{School of Natural Sciences, Institute for Advanced Study, Princeton, New Jersey 08540, USA}
\def\cern{Theoretical Physics Division, CERN, CH1211 Gen\`{e}ve 23, Switzerland}

\documentclass[prd,twocolumn,showpacs,floatfix]{revtex4}
\usepackage{graphicx}
\usepackage{dcolumn}

\begin{document}
\preprint{UFIFT-HEP-06-6}

\title{A High Resolution Search for Dark-Matter Axions}
\date{March 3, 2006}

\author{L.~D.~Duffy}
\affiliation{Department of Physics, University of Florida, Gainesville, Florida 32611, USA}
\author{P.~Sikivie}
\altaffiliation[On leave during Fall 2005 at \ias, and during Spring 2006 at ]{\cern}
\affiliation{Department of Physics, University of Florida, Gainesville, Florida 32611, USA}
\author{D.~B.~Tanner}
\affiliation{Department of Physics, University of Florida, Gainesville, Florida 32611, USA}
\author{S.~J.~Asztalos}
\affiliation{Lawrence Livermore National Laboratory, Livermore, California 94550, USA}
\author{C.~Hagmann}
\affiliation{Lawrence Livermore National Laboratory, Livermore, California 94550, USA}
\author{D.~Kinion}
\affiliation{Lawrence Livermore National Laboratory, Livermore, California 94550, USA}
\author{L.~J~Rosenberg}
\affiliation{Lawrence Livermore National Laboratory, Livermore, California 94550, USA}
\author{K.~van Bibber}
\affiliation{Lawrence Livermore National Laboratory, Livermore, California 94550, USA}
\author{D.~B.~Yu}
\altaffiliation[Present address: ]{\danny}
\affiliation{Lawrence Livermore National Laboratory, Livermore, California 94550, USA}
\author{R.~F.~Bradley}
\affiliation{National Radio Astronomy Observatory, Charlottesville, Virginia 22903, USA}

\begin{abstract}
We have performed a high resolution search for galactic halo axions in 
cold flows using a microwave cavity detector.
The analysis procedure and other details of this search are described.
No axion signal was found in 
the mass range 1.98--2.17~$\mu$eV.  We place upper limits on the density of axions in local discrete
flows based on this result.  
\end{abstract}


\maketitle

\section{Introduction}
In the current concordance cosmology, 23\% of the universe's total energy 
density is contributed by exotic dark matter \cite{WMAP}.  The axion, arising 
from the Peccei-Quinn solution to the strong CP problem \cite{PQWW}, 
satisfies the two criteria necessary for cold dark matter \cite{realign}:  (1) a 
very cold population of axions could be present in our universe
in sufficient quantities to provide the required dark matter energy density
and (2) axions are effectively collisionless; i.e.,\ their only significant 
long-range interaction is gravitational.
The Axion Dark Matter 
eXperiment (ADMX) \cite{ADMX} uses a Sikivie microwave cavity detector \cite{detect} 
to search for axions in our galactic halo.

The power in an axion signal observed by a microwave cavity detector 
is proportional to the local axion density.  
The signal width is caused by the
velocity dispersion of dark-matter axions. Therefore, in searching for axions 
it is necessary to make
some assumptions about their velocity distribution in our galactic halo.  A variety of 
galactic halo models have been put forward:  the isothermal model, results
from N-body simulations \cite{nbody} and the caustic ring model 
\cite{cau1, cau2}. 
The predictions of these models are
used to guide ADMX's search.

In the isothermal model it is expected that a significant fraction of the dark matter halo will have
an isothermal velocity distribution 
resulting from a period of ``violent
relaxation'' of the early galaxy \cite{violentrelax}.  This component of the
halo will have velocities described by a Maxwell-Boltzmann distribution.  
The ``medium 
resolution'' (MR) channel \cite{MR} searches for these axions, assuming that 
the velocity dispersion is $\mathcal{O}(10^{-3}c)$ or less.  (The escape 
velocity from our galaxy for axions is approximately $2\times10^{-3}c$.)

N-body simulations and the caustic ring model both predict substructure within
halos.  Numerical simulations indicate that hundreds
of smaller clumps, or subhalos, exist within the larger
halo \cite{nbody}.  Tidal disruption of these subhalos leads to flows in the form of
``tidal tails'' or
``streams''.  The Earth may currently be in a stream of dark matter from 
the Sagittarius A dwarf galaxy \cite{streams}.  

Non-thermalized flows from late infall
of dark matter onto the halo are also expected \cite{ips}.  
Insufficient time has elapsed for 
dark matter that has fallen into the gravitational potential of the galaxy after
violent relaxation to thermalize with the rest of the
halo.  
Matter which has fallen onto the galaxy only recently will be present in
the halo in the form of  
discrete flows.
There will be one flow of particles falling into the gravitational potential
for the first time, one flow of particles falling out for the first time, one due
to particles falling in for the second time, etc.  
Furthermore, where the gradient of the particle velocity diverges, particles 
``pile up'' and form caustics.  In the limit of zero flow 
velocity dispersion, caustics have infinite particle density.
The velocity dispersion of cold axions at
a time, $t$, prior to galaxy
formation is approximately $\delta v_a \sim 
3\times10^{-17}(10^{-5}\;\mathrm{eV}/m_a)
(t_0/t)^{2/3}$ \cite{cau2}, where $t_0$ is the present age of the universe and $m_a$
is the axion mass, constrained to lie between $10^{-6}$ and $10^{-2}$ eV 
by cosmology and astrophysical processes \cite{axrev}. 
Thus, a flow of dark matter axions will have a small velocity dispersion, 
leading to large, but finite density at the location of a caustic.

The caustic ring model predicts that the Earth is located near a 
caustic feature \cite{MW}.  Fitting the model to bumps in the Milky Way rotation curve
and a triangular feature seen in the IRAS maps predicts that the flows falling
in and out of the halo for the fifth time contain  
a significant fraction of the local halo density.  The predicted densities are
 $1.7\times10^{-24}$ g/cm$^3$ and $1.5\times10^{-25}$~g/cm$^3$ \cite{MW}, 
comparable to the local dark matter density of $9.2\times10^{-25}$ g/cm$^3$
predicted in  
\cite{density}.  The flow of the greatest density is referred to as the 
``Big Flow''.   
The possible existence of discrete flows, or streams, provides an opportunity 
to increase the ADMX discovery potential.  A discrete axion flow produces 
a narrow peak in the spectrum of microwave photons in the experiment and such
a peak can be searched for with higher signal-to-noise than a signal from
axions in an isothermal model halo.  The ``high
resolution'' (HR) channel was built to take advantage of this opportunity.
Furthermore, if a signal is found, the HR channel will provide us with 
detailed information on the structure of the Milky Way halo.

The HR channel is the most recent addition to ADMX, implemented as a simple 
addition to the receiver chain, running in parallel with the 
MR channel.  This
channel and the possible existence of discrete flows can improve
ADMX's sensitivity by a factor of 3 \cite{HR}, 
significantly enhancing its discovery potential.  The full ADMX detector is 
described in Section~\ref{sec:experiment}.    
Each discrete flow of cold axions with small velocity dispersion will be seen
as a narrow peak in the detector's output spectrum.  Our expectations for 
a signal are discussed in Section~\ref{sec:analysis}, which also contains 
the details of the 
HR analysis, the primary topic of this paper.  
After
a full search of the frequency range 478--525 MHz, no axion signals were
found and we place limits on the density of cold axions in discrete flows
in Section~\ref{sec:results}. 
This limit is compared to our previous results for the MR channel and halo
substructure predictions in Section~\ref{sec:discuss}. 

\section{Axion Dark Matter eXperiment}
\label{sec:experiment}

ADMX uses a microwave cavity detector to search for axions
in our galactic halo.  We outline the principle of the detector and briefly 
describe ADMX.  Further details of the experiment can be found in 
\cite{ADMX, Peng:2000hd}.

The microwave cavity detector utilizes the axion-electromagnetic
coupling to induce resonant conversion of axions to photons.  
The relevant interaction is
\begin{equation}
\mathcal{L}_{a\gamma\gamma}=g_{\gamma} \frac{\alpha}{\pi}
\frac{a(x)}{f_a} \mathbf{E}\cdot\mathbf{B} \; ,
\end{equation}
where $\mathbf{E}$ and $\mathbf{B}$ are the electric and magnetic fields,
$\alpha$ is the fine structure constant, $f_a$ is the axion decay constant, $a(x)$ is the axion field and 
$g_{\gamma}$ is a model-dependent coupling, of order one. 
In the Kim-Shifman-Vainshtein-Zakharov (KSVZ) model \cite{KSVZ},
$g_{\gamma} = -0.97$, and
in the Dine-Fischler-Srednicki-Zhitnitsky (DFSZ) model 
\cite{DFSZ}, $g_{\gamma} = 0.36$.
Axions in the galactic halo are non-relativistic, i.e. the energy of a single 
axion with mass $m_a$ and velocity $v$ is
\begin{equation}
E_a=m_ac^2+\frac12m_a v^2 \; ,
\label{eq:NRE}
\end{equation}
where $c$ is the speed of light.
The axion-to-photon conversion process conserves energy, i.e.\ an axion of
energy $E_{a}$ converts to a photon of frequency $\nu=E_{a}/h$.  When $\nu$ falls within the bandwidth of a cavity
mode, the conversion process is resonantly enhanced.
The signal is a peak in the spectrum 
output by the detector.
The power, $P$, developed in the cavity due to resonant axion-photon conversion
is \cite{detect}
\begin{equation}
\label{eqn-convpower}
P=\left(\frac{\alpha g_{\gamma}}{\pi f_{a}}\right)^{2}
\frac{VB_{0}^{2}\rho_{a}C}{m_{a}}\min(Q,Q_{a})\; ,
\end{equation}
where 
$V$ is the cavity volume, $B_0$ is 
the magnetic field strength, 
$\rho_{a}$ is the local density of axions with energy corresponding to the
cavity frequency, 
$Q$ is the loaded quality factor of the cavity,  
$Q_{a}$ is the ratio of the energy of the halo axions to their energy spread,
equivalent to a ``quality factor'' for the halo axion signal,
and $C$ is a mode dependent form factor which is largest for the fundamental
transverse magnetic mode, $TM_{010}$. 
The quantity $C$ is
given by
\begin{equation}
C=\frac{\left|\int_{V}d^{3}x\mathbf{E_{\omega}}\cdot\mathbf{B_{0}}\right|^{2}}
{B_{0}^{2}V\int_{V}d^{3}x\epsilon|\mathbf{E_{\omega}}|^2}\mathrm{,}
\end{equation}
in which $\mathbf{E_{\omega}}(\mathbf{x})e^{i\omega t}$ is the time dependent
electric field of the mode under consideration, $\mathbf{B_{0}}(\mathbf{x})$ is
the static magnetic field in the cavity and $\epsilon$ is the
dielectric constant of the medium inside the cavity.
The frequency-dependent form factor is evaluated numerically.
Eq.~(\ref{eqn-convpower}) can be recast in the convenient form
\begin{eqnarray}
\label{e:power}
P=&0.5\times10^{-21}\;\mathrm{W}\left({V\over 500\;\mathrm{L}}\right)
        \left({B_{0}\over 7\;\mathrm{T}}\right)^{2}C
        \left({g_{\gamma}\over 0.36}\right)^{2} \nonumber \\
	&\times\left({\rho_{a}\over 0.5\times10^{-24}\;
        \mathrm{g.cm}^{-3}}\right) 
\left({\nu_{a}\over 1 \mathrm{GHz}}
        \right)\left({\mathrm{min}[Q,Q_{a}]\over 10^5}\right)\, ,
\end{eqnarray}
where $\nu_a$ is the axion mass frequency.
As the experiment
operates with the cavity at critical coupling, half the power developed in the
cavity
is lost to its walls and half is
passed to the receiver chain.

\begin{figure*}
\centering
\resizebox{!}{0.27\textheight}{\includegraphics{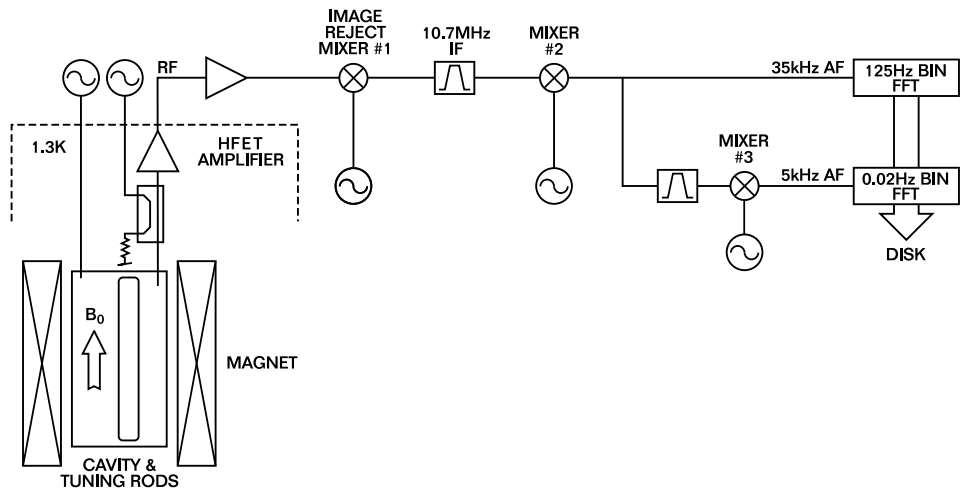}}
\caption{Schematic diagram of the receiver chain.}
\label{fig-receiver}
\end{figure*}

\begin{figure}
\centering
\resizebox{!}{0.37\textheight}{\includegraphics{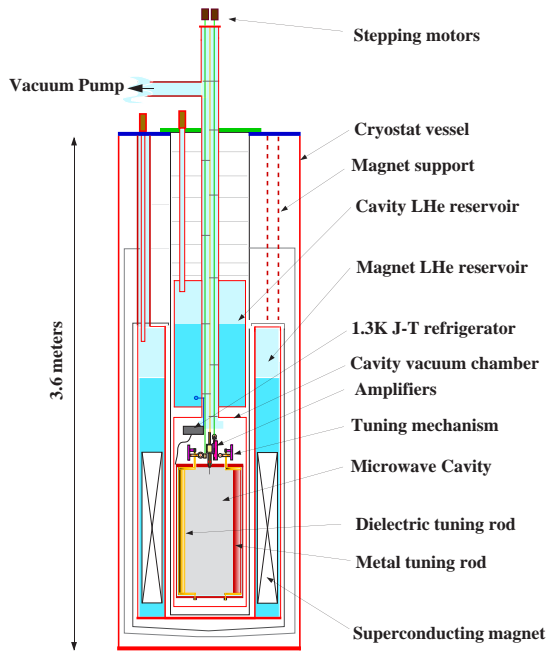}}
\caption{Sketch of the ADMX detector.}
\label{fig-detector}
\end{figure}

The HR channel became fully operational in August, 2002.
A schematic of ADMX, showing both the MR and HR channels, is given in Fig.~\ref{fig-receiver}.  A more detailed illustration of the 
magnet, cavity and cryogenic components
is shown in Fig.~\ref{fig-detector}.

The microwave cavity has an inner volume,
$V$, of 189~L.
The frequency of the $TM_{010}$ mode
can be tuned by moving a pair of metal or 
dielectric rods inside the cavity.
The cavity is located in the bore of a superconducting solenoid,
which generates a static magnetic field, $B_0$, of 7.8 T.  The voltage developed across a probe coupled to the electromagnetic field
inside the cavity is passed to the receiver chain.  
During typical operation,
$Q$ is approximately $7\times10^{4}$ and the
total noise temperature for the experiment, $T_{n}$,
is conservatively estimated to be 3.7~K, including contributions from both the
cavity and the receiver chain.

The first segment of the receiver chain is common to both the MR and HR 
channels.  It consists of a cryogenic GaAs HFET amplifier built by
NRAO, a crystal bandpass filter 
and mixers.  At the end of this segment, the signal is centered at 35~kHz, with
a 50 kHz span.  The MR signal is sampled directly after this part of the
receiver chain.  The HR channel contains an additional bandpass filter and 
mixer, resulting in a spectrum centered at 5~kHz with a 6~kHz span.

Time traces of the voltage output from the HR channel,
consisting of $2^{20}$ data points,
are taken with a 
sampling frequency of $20$~kHz. This results in a data stream of 52.4~s in length, 
corresponding to 0.019~Hz resolution in the frequency spectrum.
The data were primarily taken in parallel with the operations of the MR
channel over a period
beginning in November, 2002 and ending May, 2004.
Continuous HR coverage has been obtained and candidate peak elimination
performed for the frequency range
478--525 MHz.
Data with $Q$
less than 40 000 and/or cavity temperature above 5 K were
discarded.  In these cases, 
additional data were taken to ensure coverage of the entire range.

\section{High resolution analysis}
\label{sec:analysis}

We use the HR channel to search for narrow peaks caused by flows of cold 
axions through the detector.  The background is
thermal and electronic noise plus narrow lines from the rf environment
of the experiment.  These environmental peaks are signals from nearby emitters (e.g.\ 
computer clocks) which leak into the cavity by a variety of means.  

When placing limits on cold flows of axions, we assume that the flows are 
steady, i.e.\ the rates of change of velocity, velocity dispersion and density
of the flows are slow compared to the time scale of the experiment.
The assumption of a steady flow implies that the signal we are searching
for is always present.  Even so, the kinetic energy term in Eq.(\ref{eq:NRE}) 
and the corresponding signal frequency change over time due to the Earth's 
rotational and orbital motions. 
In addition to a signal frequency shift in data taken at different times, 
apparent broadening
of the signal occurs because its frequency shifts while the data are being taken.

In this section, we first describe the signal frequency shifts due to the 
Earth's rotation and orbital motion, and the associated signal broadening
(subsection~A).  Next, we describe the properties of the
noise in the HR channel (subsection~B).  In 
subsection~C, we describe how we correct each spectrum for
systematic effects introduced by the receiver chain.  Finally, in 
subsection~D, we describe how candidate peaks are selected
and shown not to be caused by cold flows of dark matter axions.

\subsection{Axion signal properties}
\label{ss:sigprop} 
An axion signal will undergo diurnal and annual modulations due to 
the Earth's rotation and orbital motion, respectively \cite{Ling:2004aj}.
Thus, the frequency
at which axions are resonantly converted to photons will shift.  
We show here that this frequency modulation cannot move a signal by an 
amount which is larger than the detector bandwidth.
As the energy of an axion in the ADMX detector is non-relativistic 
(Eq.~(\ref{eq:NRE})), the shift in frequency of the signal,
$\Delta f$, due to a change in velocity of the axion flow relative to the 
detector, $\Delta v$, is
\begin{equation}
\Delta f = \frac{f v \Delta v}{c^2} \; .
\label{eq:sigmove}
\end{equation}
We have investigated the magnitude of both annual and diurnal signal 
modulation at $f=$ 500~MHz.  The velocity of a dark matter flow relative to the Earth will
be in the range 100--1000 km/s.  We chose 600 km/s as a representative value
for the purpose of estimation.
For the daily modulation, we have assumed that the
detector is located at the Earth's equator and that, in the frame in which the axis
of rotation of the Earth is stationary, the flow velocity is first aligned and 
then anti-aligned (or vice versa) with the detector's motion due to the Earth's rotation over the
course of a day.  These assumptions result in the largest possible change in relative velocity
between the detector and the flow due to the Earth's rotation.  The Earth's
rotational velocity is 0.4 km/s at the equator. 
The resulting daily signal modulation is of order 1~Hz.
For the annual modulation, to again maximize the change in relative velocity,
we have also considered the case of extreme flow velocity alignment with the
Earth's orbital motion.  The Earth's orbital velocity is 30 km/s.
The annual modulation
produces the larger frequency shift,
of order 100~Hz within a year.  
The HR channel has a 6~kHz bandwidth, ensuring that an axion
peak will appear in spectra taken with center frequency equal to a previously
observed axion signal frequency.

The signal broadening, $\delta f$, due to a change $\delta v$ in flow velocity
while data are being taken is
\begin{equation}
\delta f = \frac{f v \delta v}{c^2} \; .
\label{eq:sigbroad}
\end{equation}
The most significant signal broadening is due to the Earth's rotation.
Using the same assumptions as before, we find that the broadening
is at most $4\times10^{-3}$ Hz during the 52 s taken to acquire a single
time trace.  This is less than the spectral resolution of 0.019 Hz.
The broadening due to the Earth's orbital motion is only of order 
$10^{-4}$~Hz in this same time interval.

As the signal broadening due to the Earth's rotation and orbital motion is
negligible, we can use Eq.~(\ref{eq:sigbroad}) to relate the width, 
$\delta f$, of a signal peak to the velocity dispersion, $\delta v$, of the
axion flow that causes it.  In general, we do not know the velocity dispersion
of the cold axion flows which we search for, although we note that
\cite{MW} claims an upper limit of 53 m/s on the velocity dispersion
of the Big Flow.  Subsequently, we do not know the signal width.  
To compensate, we perform our search at multiple resolutions by combining 0.019~Hz wide
bins. 
These searches are referred to as $n$-bin searches,
where $n =$ 1, 2, 4, 8, 64, 512 and 4096.  For $f=500$ MHz and $v=300$~km/s, 
the corresponding flow velocity dispersions are
\begin{equation}
\delta v_n= 12\,n \;\mathrm{m}/\mathrm{s}\left(\frac{300 \mathrm{km}/
	\mathrm{s}}{v}\right) \; .
\label{eq:vdisp}
\end{equation}
Further details on the $n$-bin searhces are given in Section~\ref{ss:sproc}.
                                                                                
\subsection{Noise in the HR Channel}
\label{ss:hrnoise}
The power output from the HR channel is expressed in units of $\sigma$,
the rms noise power.
This noise power is related to the noise temperature, $T_n$, via 
\begin{equation}
\sigma=k_B T_n \sqrt{\frac{b}{t}} \; ,
\end{equation}
where $k_B$ is Boltzmann's constant, $b$ is the frequency resolution 
and $t$ is the acquisition time.  The total noise
temperature $T_n=T_{C}+T_{el}$, where $T_{C}$ is the physical cavity
temperature and $T_{el}$ is the electronic noise contribution from the
receiver chain.
As no averaging is performed in HR
sampling, $b=1/t$.  Thus, the rms noise power is
\begin{equation}
\sigma=k_B b T_n \; .
\label{eq:Pnoise}
\end{equation}
Output power is normalized to $\sigma$ and $T_n$ is
used to determine this power.  
We verified Eq.~(\ref{eq:Pnoise}) experimentally by allowing
the cavity to warm and observing that $\sigma$ is proportional to 
$T_{C}$.  As this is our calibration of the power output from the cavity,
it is important that we understand the noise in the HR channel.

The noise in the HR channel is observed to have an exponential 
distribution.  We now explain why this is expected.    
The noise in a single bin
is the sum of independent sine and cosine components, as no averaging 
occurs.  
We expect that the noise amplitude, $a$, for a
single component
(i.e. sine or cosine) has a Gaussian probability distribution, 
\begin{equation}
\label{eqn-gauss}
\frac{dP}{da}=\frac{1}{\sqrt{2\pi}\sigma_{a}}\exp\left(-\frac{a^{2}}{2
\sigma_{a}^{2}}\right) \, ,
\end{equation}
where $\sigma_{a}$ is the standard deviation.
Indeed, the energy distribution should be proportional to a Boltzmann factor,
$\exp(-E/kT)$, and
non-relativistic and classical energies, such as $E=mv^{2}/2$ or
$E=kx^{2}/2$ are proportional to squares of the amplitude.

As there are two components
per bin, the addition of $n$ bins is that of $2n$ independent
contributions.  
The sum of $2n$ independent normal-distributed components is described by
a chi-square distribution with $2n$ degrees of freedom.  Thus, the probability
distribution for an $n$-bin is a $\chi^2(2n)$ distribution.  We demonstrate
this explicitly in the following.

The probability distribution, $dP/dp_n$, of observing noise
power $p_n$ in an $n$-bin is 
\begin{equation}
\frac{dP}{dp_n}=\left(\prod_{i=1}^{2n}\int^{\infty}_{-\infty}da_i\right)
	\frac{\exp(-\frac{1}{2\sigma_a^2}\sum_{j=1}^{2n} a_j^2)}{(\sqrt{2\pi}
		\sigma_a)^{2n}}\,\delta(p_n-\sum_{k=1}^{2n}\frac{a_k^2}2
		)\, .
\end{equation}
Evaluating the above expression,
\begin{equation}
\frac{dP}{dp_{n}}=\frac{{p_{n}}^{n-1}}{(n-1)!\sigma_{a}^{2n}}
        \exp\left(-\frac{p_{n}}{\sigma_{a}^2}\right) \; .
\label{eq:nprob}
\end{equation}
For $n=1$,
\begin{equation}
\label{eqn-sigaprob}
\frac{dP}{dp_{1}}=\frac{1}{\sigma_{a}^{2}}\exp\left(-\frac{p_{1}}{\sigma_{a}^{2}}\right)
        \mathrm{,}
\end{equation}
which is indeed a simple exponential, as expected.

Using this noise distribution, we can easily see that the average (rms) noise
power in the one bin search, $\sigma$, is $\sigma={\sigma_{a}}^2$.
Substituting this in Eq.~(\ref{eqn-sigaprob}), the noise power
distribution function becomes
\begin{equation}
\label{eqn-npprob}
\frac{dP}{dp_{1}}=\frac{1}{\sigma}
        \exp\left(-\frac{p_{1}}{\sigma}\right) \; .
\end{equation}

\begin{figure}
\centering
\includegraphics{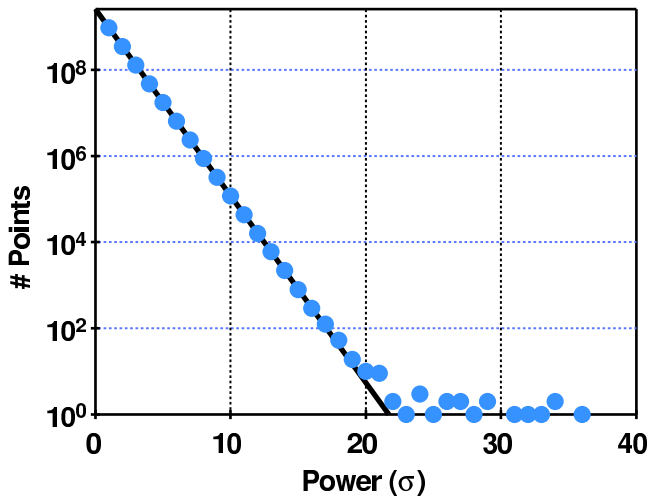}
\caption{Power distribution for a large sample of 1-bin data.}  
\label{fig:stat1}
\end{figure}

For each individual spectrum, the baseline noise level, $\sigma$, is determined
by plotting the number of frequency bins, $N_{p}$, with power
between $p$ and $p+\Delta p$ against $p$.  According to Eq.~(\ref{eqn-npprob}),
\begin{equation}
N_{p}=\frac{N\Delta p}{\sigma}\exp\left(-\frac{p}{\sigma}\right) \; ,
\end{equation}
where $N$ is the total number of data points.  As
\begin{equation}
\ln N_{p}=-\frac{p}{\sigma}+\ln\left(\frac{N\Delta p}{\sigma}\right) \; ,
\label{eqn-1bindist}
\end{equation}
$\sigma$ is the inverse of the slope of the $\ln N_{p}$ versus $p$ plot.
Fig.~\ref{fig:stat1} demonstrates that the data is in good agreement
with this relation for $p$ less than $20\sigma$.  The deviation of the data
from Eq.~(\ref{eqn-1bindist}) for $p$ greater than $20\sigma$ is due to the fact
that our background is not pure noise, but also contains environmental signals
of a non-statistical nature.
                
\begin{figure}
\centering
\includegraphics{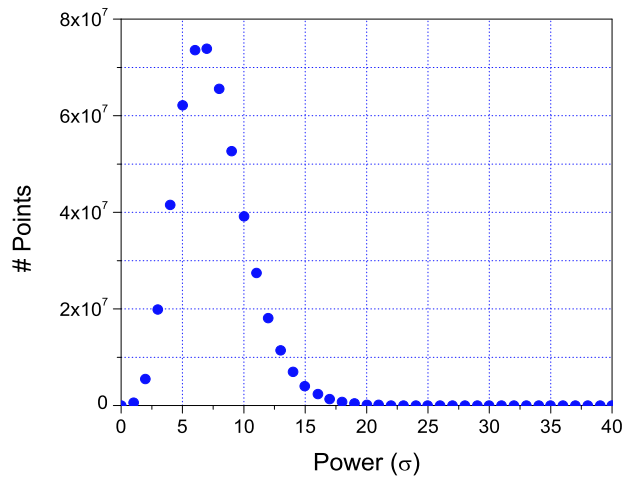}
\caption{Power distribution for a large sample of 8-bin data.}
\label{fig:8stats}
\end{figure}

\begin{figure}
\centering
\includegraphics{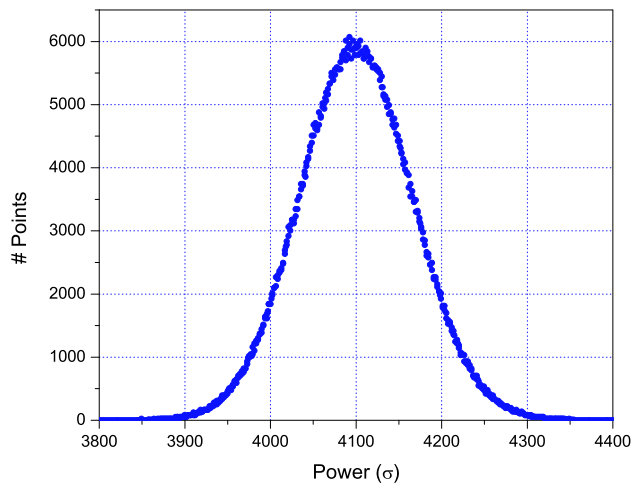}
\caption{Power distribution for a large sample of 4096-bin data.}
\label{fig:nstats}
\end{figure}

As we combine an increasing number of bins, the noise power probability 
distribution approaches a Gaussian, in accordance with the central limit 
theorem.  The right-hand side of Eq.~(\ref{eq:nprob}) approaches a Gaussian in the
limit of large $n$.  We have examined a large sample of noise in each
$n$-bin search and verified that it is distributed according to 
Eq.~(\ref{eq:nprob}).  Figure~\ref{fig:8stats} illustrates statistics for the 
8-bin search and Fig.~\ref{fig:nstats} shows statistics for the
4096-bin search, a near Gaussian curve.  Figures~\ref{fig:stat1} through 
\ref{fig:nstats} illustrates the progression from exponential to Gaussian
noise power probability distribution.

\begin{figure}
\includegraphics{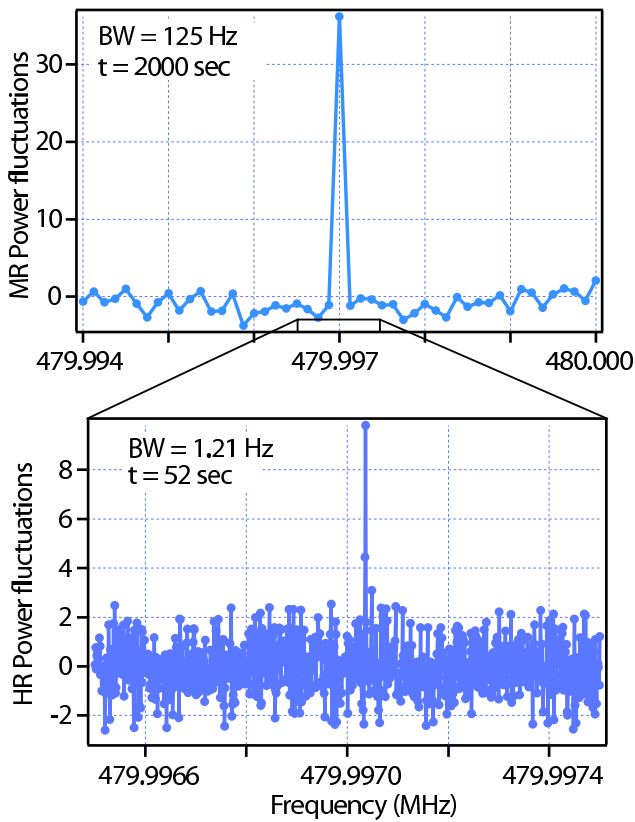}
\caption{An environmental peak as it appears in the MR search (top) and the 
64--bin HR search.  The unit for the vertical axis is the rms power 
fluctuation in each case.}
\label{fig:crosscal}
\end{figure}

In addition to examining the behavior of the noise statistics, we have 
performed a cross-calibration between the HR and MR channels.  
The signal
power of an environmental peak, 
observed at 480~MHz and shown in Fig.~\ref{fig:crosscal}, was examined in both the HR and MR channels.  
The observed HR signal power was $(1.8 \pm 0.1)\times10^{-22}$~W, where the
error quoted is the statistical uncertainty.  
The MR channel
observed signal power $1.7\times10^{-22}$~W, in agreement with the
HR channel.
Note that the MR signal was acquired
with a much longer integration time than that of the HR signal (2000~s for 
MR versus 52~s for HR).  

The combination of the calibration of the noise power with cavity temperature, the 
consistency between expected and observed noise statistics, and the agreement of 
signal power observed in both the HR and MR channels, makes us confident that
the signal power is accurately determined in the HR channel.

\subsection{Removal of systematic effects}
\label{ss:system}
There are two systematic effects introduced in the receiver
chain shown in Fig.~\ref{fig-receiver}.  Two passband filters are present on the HR receiver chain:  one
with bandwidth 35~kHz on the shared MR-HR section and a passive LC filter 
of bandwidth
6~kHz, seen by the HR channel only.  The combined response of both these 
filters has been analyzed and removed from the data.  The second systematic
effect is due to the frequency-dependent response of the coupling between
the cavity and the first cryogenic amplifier. This effect is
removed using the equivalent
circuit model described later.

\begin{figure}
\centering
\includegraphics{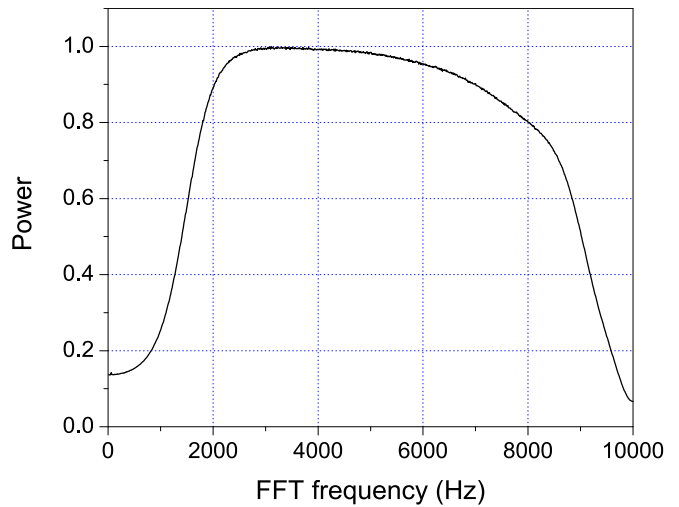}
\caption{HR filter response calibration data (512 bin average).  
The power has been normalized to the maximum power output.}
\label{fig-filtercal}
\end{figure}

The combined passband filter response was determined by taking data with 
a white noise source at the rf input of the receiver chain.  A total of 
872 time traces were recorded over a two day period.  In order to
achieve a reasonably smooth calibration curve, 512 bins in the 
frequency spectrum for each time trace were averaged giving  
9.77 Hz resolution.
The combined average of all data is shown in Fig.~\ref{fig-filtercal}.  This 
measured response was removed from all data used in the HR search, as follows.
The raw
power spectra have frequency 0--10~kHz, where the center frequency of 5~kHz 
has been mixed down from the cavity frequency.  Each raw power spectrum is
cropped to the region 2--8~kHz to remove the frequencies not within the LC filter
bandwidth.   Each remaining frequency bin is then weighted by a factor
equal to the receiver chain response at the given frequency divided by the
maximum receiver chain response. 
Interpolation for frequency points not specifically included in the calibration
curve is performed by assuming that each point on the calibration curve
was representative of 512 bins centered on that frequency, so all power
corresponding to frequencies within that range is normalized by the same
factor.  As the calibration curve varies slowly with frequency within the
window to which each spectra is cropped, this is an adequate
treatment of the normalization.  
                                                                                
\begin{figure}
\centering
\includegraphics[width=0.45\textwidth, bb=18 14 100 84]{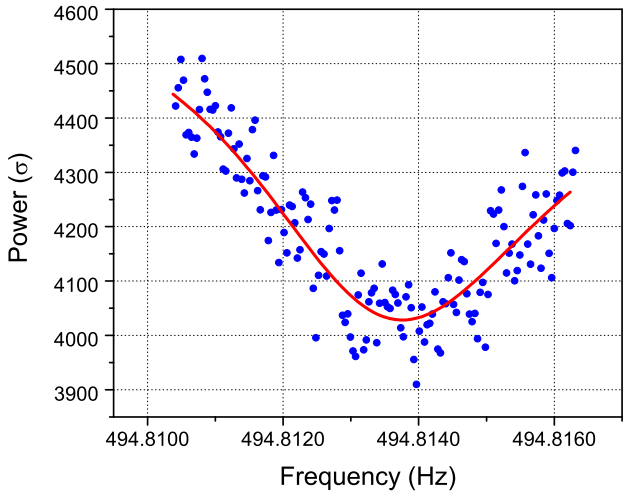}
\caption{Sample 4096-bin spectrum before 
correction for the cavity-amplifier coupling.
The line is the fit obtained using the equivalent
circuit model.}
\label{fig-4096before}
\end{figure}

In the MR channel, the effect of the cavity-amplifier coupling is 
described using an equivalent-circuit model \cite{DawPhD}.  This model has
been adapted for use in the HR channel.  The frequency dependent response
of the cavity amplifier coupling is most evident in the 4096-bin search, thus
this is the data used to apply the equivalent circuit model.  A sample 
spectrum before correction is shown in Fig.~\ref{fig-4096before}.  

In the equivalent-circuit model, each frequency is given by $\Delta$, the
number of bins it
is offset
from the bin of the center frequency, measured in units of the
4096-bin resolution, i.e.\ $b_{4096}=78.1$~Hz.
The equivalent-circuit model predicts that
the power (in units of the rms noise)
at the NRAO amplifier output  (the point labelled ``RF'' in
Fig.~\ref{fig-receiver}) in the 4096--bin search at the frequency offset
$\Delta$ is
\begin{equation}
P(\Delta)=\frac{a_{1}+8a_{3}\left(\frac{\Delta-a_{5}}{a_{2}}\right)^{2}
+4a_{4}\left(\frac{\Delta-a_{5}}{a_{2}}\right)}{1+4\left(\frac{\Delta-a_{5}}{a_{2}}\right)^{2}}
\mathrm{,}
\label{eq:5fit}
\end{equation}
where the parameters $a_{1}$ through $a_{5}$ are
\begin{eqnarray}
a_{1} & = & (b_{4096}/b)(T_{C} + T_{I} + T_{V})/T_{n}\; , \\
a_{2} & = & f_{0}/(b_{4096}\,Q)\; , \\
a_{3} & = & (b_{4096}/b)(T_{I} + T_{V} + (T_{I} - T_{V})\cos(2kL))/T_{n} \; ,
         \\
a_{4} & = & (b_{4096}/b)((T_{I} - T_{V})\sin(2kL))/T_{n}\;\mathrm{and} \\
a_{5} & = & (f_{0} - f_{cen})/b_{4096} \; . 
\end{eqnarray}
In the above expressions, $T_{C}$ is the physical temperature of the microwave cavity, $T_{I}$
and $T_{V}$ are the current and voltage noise, respectively, contributed by
the amplifier, $T_{n}$ is the noise temperature contributed from all 
components, $b$ is the frequency resolution of the HR channel, 
i.e.\ 0.019 Hz,
$L$ is the electrical (cable) length from the cavity to the HFET
amplifer, $f_{0}$ is the cavity resonant frequency, $f_{cen}$ is the center
frequency of the spectrum and $k$ is the wavenumber
corresponding to frequency $f_{cen}+b\Delta$.  The factor $b_{4096}/b$ appears
in the parameters $a_{1}$, $a_{3}$ and $a_{4}$ as it is an overall factor
which results from normalizing the power to the single bin noise baseline.
In practice, the parameters $a_{1}$ through $a_{5}$ are established 
by fitting.
The line in Fig.~\ref{fig-4096before}
shows the fit obtained using the equivalent circuit model.

Large peaks in the data, e.g.\ an axion signal or environmental peak, are 
removed before fitting to prevent bias.  The 4096-bin spectrum is used to
perform the fit and then the original 1-bin spectrum is corrected to
remove the systematic effect.  The weighting factors are calculated using
Eq.~(\ref{eq:5fit}) and the fitted parameters, $a_1$ through $a_5$, at the
center of each bin of width $b_{4096}$.  These factors are 
the ratio of the fit at a given point 
to the maximum value of the fit.  Each 1-bin is multiplied by
the factor calculated for the bin of width $b_{4096}$ within which it falls.

The removal of the cavity-amplifier coupling and the passband filter
response using the techniques described above has been demonstrated to result
in flat HR spectra.

\subsection{Axion signal search procedure}
\label{ss:sproc}

We now describe the search for an axion signal and summarize the analysis
performed on each time trace.  

The width of an axion signal is determined by the signal frequency, axion
velocity and flow velocity dispersion (Eq.~(\ref{eq:sigbroad})), the latter being the most uncertain
variable.
$n$-bin searches, where $n$ is the number of 
adjacent 1-bins added together ($n=$ 1, 2,
4, 8, 64, 512 and 4096), are conducted to allow for various velocity 
dispersions.  For searches with $n>1$, 
there is an overlap between successive $n$-bins such that each $n$-bin 
overlaps with the last half of the previous and first half of the following
$n$-bin.
This scheme is illustrated
for the 2, 4 and 8-bin searches in Fig.~\ref{f:coadd}.
\begin{figure*}
\begin{tabular}{||lc||}
\hline \hline
1--bin search:& \fbox{1} \fbox{2} \fbox{3} \fbox{4} \fbox{5}
        \fbox{6} \fbox{7} \fbox{8} \fbox{9} \fbox{10} \fbox{11} \fbox{12}
        \fbox{13} \fbox{14} \fbox{15} \fbox{16} \\
\hline
2--bin search:&\fbox{1 2} 
        \fbox{3 4} 
        \fbox{5 6}
        \fbox{7 8}
        \fbox{9 10} 
        \fbox{11 12} 
        \fbox{13 14} 
        \fbox{15 16} \\
& \fbox{2 3} \fbox{4 5} \fbox{6 7} \fbox{8 9} \fbox{10 11} \fbox{12 13}
        \fbox{14 15}\\
\hline
4--bin search:& \fbox{1 2 3 4}  \fbox{5 6 7 8}
        \fbox{9 10 11 12}  \fbox{13 14 15 16} \\
&
\fbox{3 4 5 6}
\fbox{7 8 9 10} \fbox{11 12 13 14}\\
\hline
8--bin search:&\fbox{1 2 3 4 5 6 7 8}
        \fbox{9 10 11 12 13 14 15 16}\\
&\fbox{5 6 7 8 9 10 11 12}\\
\hline \hline
\end{tabular}
\caption{Illustration of the addition scheme for the 2, 4 and 8-bin searches.
The numbers correspond to the data points of the 1-bin search.  Numbers within
the same box are bins added together to form a single datum in the 
$n$-bin searches with $n>1$.}
\label{f:coadd}
\end{figure*}
                                                   
The search for an axion signal is performed by scanning each spectrum for
peaks above a certain threshold.  All such peaks are considered candidate
axion signals.  
The thresholds are set at a
level where there is only a small probability that a pure noise peak will
occur and such that the number of frequencies considered
as candidate axion peaks is manageable.
The candidate thresholds used were 20, 25, 30, 40, 120, 650 and 4500 $\sigma$, 
in increasing order of $n$.

All time traces are analyzed in the same manner.  
A fast Fourier transform is performed 
and an initial estimate of $\sigma$ is obtained by fitting the 1-bin
noise distribution to Eq.~(\ref{eqn-1bindist}).
Systematic effects are then removed, i.e.\ 
the corrections described in Section~\ref{ss:system} for the filter passband response
and cavity--amplifier coupling are performed.  ``Large'' peaks not included in
the equivalent circuit model fit for the cavity-amplifier response are
defined to be those greater than 120\% of the search threshold for each
$n$--bin search.  
After the removal of systematic effects, the 1-bin noise distribution is
again fitted to Eq.~(\ref{eqn-1bindist}) to obtain the true value of 
$\sigma$ and the search for peaks above the thresholds takes place.

The axion mass is not known, requiring that a 
range of frequencies must be examined. 
Full HR coverage has been obtained for the region 478--525 MHz, corresponding
to axion masses between 1.98 and 2.17~$\mu$eV.  
The selected frequency range is examined in three stages for axion peaks, as
follows:

\emph{Stage 1:}  Data for the entire selected frequency range is taken.
The frequency step between successive spectra is approximately 1~kHz, 
i.e.\ the center frequency of each spectrum differs from the previous spectrum
by 1~kHz.  Frequencies at which candidate axion peaks occur are recorded for
further examination during stage 2.

\emph{Stage 2:}  Multiple time traces are taken at each candidate frequency
from stage 1.  The steady flow assumption described in Section~\ref{ss:sigprop} means
that a peak will appear in spectra taken with center frequency equal
to the candidate frequency from stage 1 if such a peak is an axion signal.  
The frequencies of persistent peaks, i.e.\ peaks that appear during both
stage 1 and 2 are examined further in stage 3.

\emph{Stage 3:}  Frequencies of persistent peaks undergo 
a three-part examination.  The first
step is to repeat stage 2, to ensure the peaks still persist.
Secondly, the warm port attenuator is removed from the cavity and multiple time
traces taken.
If the peak is due to external
radio signals entering the cavity (an environmental peak), 
the signal power will increase dramatically.
If the signal originates in the cavity due to
axion-photon conversion, the power developed in the cavity
will remain the same as that for the normal configuration.  
The third step is
to use an external antenna probe as a further confirmation that the signal
is environmental.
Some difficulties were encountered with the antenna probe,  
due to polarization of environmental signals.  However, 
the second step
is adequate to confirm that peaks are environmental.  If a persistent peak
is determined to not be environmental, a final test will confirm that it
is an axion signal.  The power in such a signal must grow proportionally 
with the 
square of the magnetic field ($B_0$ in Eq.~(\ref{e:power})) and disappear
when the magnetic field is switched off.

No axion peaks were found in the range 478--525~MHz using this approach.  The
exclusion limit calculated from this data is discussed in the following
section.
                                                                                
\section{Results}
\label{sec:results}
Over the frequency range 478--525~MHz, we derive an upper limit on the
density of individual flows of axion dark matter as a function of the
velocity dispersion of the flow. 
The corresponding axion mass range is 1.97--2.17 $\mu$eV.  Each $n$-bin
search places an upper limit on the density of a flow with maximum velocity
dispersion, $\delta v_n$, as given by Eq.~(\ref{eq:vdisp}).

Several factors reduce the power developed in an axion peak from that
given in Eq.~(\ref{e:power}).  
The experiment is operated near critical coupling of the cavity to the 
preamplifier, so that
half this power 
is observed when the cavity resonance frequency, $f_0$, is
precisely tuned to the axion energy.  If $f_0$ is not at the center
of a $1$-bin, the power is spread into adjacent bins, as discussed below.
When the axion energy is off-resonance, 
but still within the cavity bandwidth at a frequency $f$, 
the Lorentzian cavity response reduces
the power developed by an additional factor of
\begin{equation}
h(f)=\frac{1}{1+4Q^2\left(\frac{f}{f_{0}}-1\right)^2}\; .
\end{equation}
To be conservative, we calculate the limits at points where successive
spectra overlap, i.e.\ at the frequency offset from $f_0$ that minimizes 
$h(f)$.

If a narrow axion peak falls at the center of a 1-bin, all power is deposited in that
1-bin.  
However, if such a peak does not fall at the center of a 1-bin, the power
will be spread over several 1-bins.  We now calculate the minimum power
in a single $n$-bin caused by a randomly situated, infinitely narrow 
axion line.  
The data recorded is the voltage
output from the cavity as a function of time.  The voltage as a function of 
frequency is obtained by Fourier transformation and then squared to obtain a
raw ``power'' spectrum.  The actual power is obtained by comparison to the rms noise 
power.  The data are sampled for a finite amount of time and thus, the Fourier transformation of the 
output, 
$\mathcal{F}(f)$,  will
be of the voltage multiplied by a windowing function, i.e.
\begin{equation}
\mathcal{F}(f)=\int_{-\infty}^{\infty}v(t)w(t)\exp(i2\pi ft)dt \; ,
\label{eq:ft1}
\end{equation}    
where $v(t)$ is the measured output voltage and $w(t)$ is the windowing 
function for a sampling period $T$,
\begin{equation}
w(t)=\left\{ \begin{array}{ll}
	1 & \mathrm{if }-T/2\leq t \leq T/2 \; ,\\
	0 & \mathrm{otherwise} \; .
	\end{array}
	\right.
\end{equation}
Eq.~(\ref{eq:ft1}) is equivalent to
\begin{equation}
\mathcal{F}(f)=\int_{-\infty}^{\infty}V(k)W(f-k)dk \; ,
\label{eq:ft2}
\end{equation}
where $V(f)$ and $W(f)$ are the Fourier transforms of the output voltage, 
$v(t)$, and the windowing function, $w(t)$, i.e.\ $\mathcal{F}(f)$ is the
convolution of $V(f)$ and $W(f)$, given by
\begin{equation}
W(f)=\frac{\sin(\pi f T)}{\pi f} \; .
\label{eq:sinc}
\end{equation} 
Discretizing Eq.~(\ref{eq:ft2}) and inserting Eq.~(\ref{eq:sinc}), we have
\begin{equation}
\mathcal{F}(f)=\sum_{m=0}^N V((m+\frac12)b) 
	\frac{\sin(\pi(\frac{f}{b}-(m+\frac12)))}
		{\pi(\frac{f}{b}-(m+\frac12))} \; ,
\label{eq:ft3}
\end{equation}
where $b$ is the frequency resolution of the HR channel, $2N$ points 
are taken in the original time trace, and 
the center frequency of the $j$th 1-bin is $(j+1/2)b$.  Thus, for an axion
signal of frequency $f$ falling in 1-bin $j$, 
a fraction of the power 
\begin{equation}
g(m)=\left(\frac{\sin(m\pi+\delta)}{m\pi+\delta}\right)^2\; ,
\label{eq:sinc2}
\end{equation}
is lost to the $m$th 1-bin from 1-bin $j$, where 
$\delta = \pi(m+1/2 - f/b)$.  If $\delta =0$, i.e\ the axion signal frequency
is exactly equal to a 1-bin center frequency, all the power is deposited in
a single 1-bin.  However, if this is not the case, power is lost to other
1-bins.  In setting limits, we assume that the power loss is maximal.

The maximum power loss occurs when a signal in the 1-bin search 
falls exactly between the center frequency of two adjacent 1-bins.  In this
case, when $\delta=\pi/2$, Eq.~(\ref{eq:sinc2}) shows that 40.5\% of the 
power will be deposited in each of two 1-bins.  In $n$-bin seaches with 
$n\geq 2$, not as much power is lost to other $n$-bins, due to the overlap
between successive $n$-bins.  The minimum power deposited in an $n$-bin is
81\% for $n=2$, 87\% for $n=4$ and 93\% for $n=8$. 
For $n=$ 64, 512 and 4096, the amount of power not deposited in a single $n$-bin
is negligible.

For the $n$-bin searches with $n=$ 64, 512 and 4096, a background noise
subtraction was performed which will lead to exclusion limits at the 
97.7\% confidence level.  
These limits are derived using the power at which the sum of the
signal power and background noise power have a 97.7\% probability to
exceed the candidate thresholds.  
We call this power the ``effective''
threshold for each search.   
The effective thresholds are obtained by integrating the noise probability distribution, Eq.~(\ref{eq:nprob}), numerically 
solving for the background noise power corresponding to the 97.7\% confidence level for each
$n$ and 
subtracting these values from the original candidate thresholds.  For
$n=$ 64, 512 and 4096, the effective thresholds are 71, 182 and 531 
$\sigma$, respectively.  For smaller values of $n$, background noise 
subtraction does not significantly improve the limits and the effective
threshold was taken to be the candidate threshold.  
Table~\ref{t:resultinfo} summarizes this information and shows the 
frequency resolution of each search with the corresponding maximum flow 
velocity dispersion from Eq.~(\ref{eq:vdisp}) for $v=600$ km/s.
\begin{table}
\caption{\label{t:resultinfo}Effective power thresholds for all $n$-bin searches, with the
frequency resolutions, $b_n$ and corresponding maximum flow velocity 
dispersions, $\delta v_n$, for a flow velocity of 600~km/s.}
\begin{ruledtabular}
\begin{tabular}{rrdr}
$n$ & Effective & b_n & $\delta v_n$ \\
 &threshold ($\sigma$) & \text{(Hz)} & (m/s) \\
\hline
1 & 20 & 0.019 & 6 \\ 
2 & 25 & 0.038 & 10 \\ 
4 & 30 & 0.076 & 20 \\ 
8 & 40 & 0.15 & 50 \\ 
64 & 71& 1.2 & 400 \\ 
512 & 182& 9.8 & 3000 \\ 
4096 & 531& 78 & 20000\\ 
\end{tabular}
\end{ruledtabular}
\end{table}

Our exclusion limits were calculated for an axion signal with power
above the effective threshold reduced by the appropriate factors.  These 
factors arise from the  
critical coupling, the Lorentzian cavity response and the maximum power loss due
to the peak not falling in the center of an $n$-bin, as outlined above.
Equations~(\ref{e:power}) and (\ref{eq:Pnoise}) were used, for both KSVZ and DFSZ 
axion couplings.  The cavity volume, $V$, is 189 L.  Measured values of
the quality factor, $Q$, the magnetic field, $B_0$, and the 
cavity temperature, $T_C$, are recorded in each data file.  Numerically 
determined values of the form factor, $C$ are given in 
Table~\ref{t:formfactor}.  The electronic noise temperature, $T_{el}$, 
was conservatively taken from the specifications of the NRAO amplifier, 
the dominant source of noise in the receiver chain, although our measurements
indicate that $T_{el}$ is less than specified.  These values are also given
in Table~\ref{t:formfactor}. 
Linear interpolation between
values at the frequencies specified was used to obtain values of $C$ and 
$T_{el}$ at all frequencies. 
\begin{table}
\caption{\label{t:formfactor}Numerically calculated values of the form 
factor, $C$, and amplifier noise temperatures, $T_{el}$, from NRAO 
specifications.}
\begin{ruledtabular}
\begin{tabular}{cdd}
Frequency (MHz) & \text{C} & T_{el} \text{(K)}\\ \hline 
450&0.43&1.9\\ 
475&0.42&1.9\\ 
500&0.41&1.9\\ 
520&0.38&1.9\\ 
550&0.36&2.0\\
\end{tabular}
\end{ruledtabular}
\end{table}

\begin{figure}
\resizebox{0.45\textwidth}{!}{\includegraphics{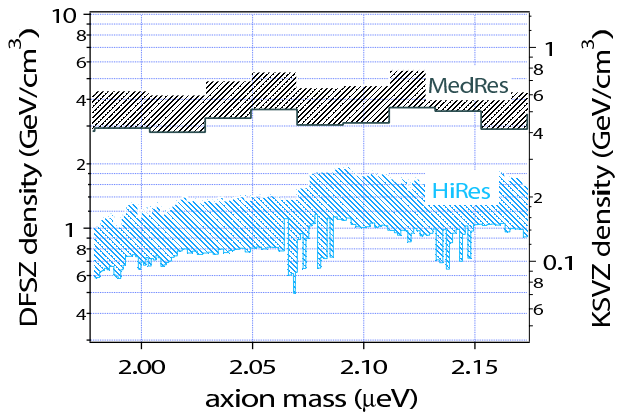}}
\caption{97.7\% confidence level limits for the HR 2-bin search on the density
of any local axion dark matter flow as a function of axion mass, for the DFSZ
and KSVZ $a\gamma\gamma$ coupling strengths.  Also shown is the previous 
ADMX limit using the MR channel.  The HR limits assume that
the flow velocity dispersion is less than $\delta v_2$ given by 
Eq.~(\ref{eq:vdisp}).}
\label{f:limits2}
\end{figure}

The 2-bin search density exclusion limit obtained using these values is shown in
Fig.~\ref{f:limits2}.  For values of $n$ other than $n=2$, the exclusion limits
differ by only constant factors.  The constant factors
are 1.60, 1.00, 1.12, 1.39, 2.53, 5.90 and 17.2 for $n=$ 1, 2, 4, 8, 64, 512
and 4096, respectively.  

\section{Discussion}
\label{sec:discuss}

We have obtained exclusion limits on the density in local flows of cold axions
over a wide range of velocity dispersions.  The most stringent limit, shown
in Fig.~\ref{f:limits2}, is from the 2-bin search.  For a flow velocity
of 600 km/s relative to the detector, the 2-bin search corresponds to a 
maximum flow velocity dispersion of 10 m/s.  The 1-bin search limit is less
general, in that the corresponding flow velocity dispersion is half that of the
2-bin limit.  It is also less stringent; much more power may be lost due
to a signal occurring away from the center of a bin than in the $n=2$ case.
For $n>2$, the limits are more general, but the larger power threshold of the 
searches make them less stringent.

The largest flow predicted by the caustic ring model has density 
$1.7\times10^{-24}$ g/cm$^3$ ($0.95$ GeV/cm$^3$), velocity of 
approximately $300$~km/s relative
to the detector, and velocity dispersion less than 53 m/s \cite{MW}.  
Using Eq.~(\ref{eq:sigbroad}) with Table~\ref{t:resultinfo} and the information displayed in 
Fig.~\ref{f:limits2} multiplied by the appropriate factors of 1.12 to obtain
the 4-bin limit, 
it can be seen that the 4-bin search, corresponding to
maximum velocity 50 m/s for $v=300$ km/s, would detect this flow if it 
consisted of  
KSVZ axions.  For DFSZ axions, this flow would be detected for approximately
half the search range. 

Figure~\ref{f:limits2} demonstrates that the high resolution analysis
improves the detection capabilities of ADMX when a significant fraction of the
local dark matter density is due to flows from the incomplete thermalization
of matter that has only recently fallen onto the halo.  The addition of this
channel to ADMX provides an improvement of a factor of 3 over our previous
medium resolution analysis.\\

\section*{Acknowledgements}
This research is supported in part by the U.S. Department of Energy under
Contract W-7405-ENG-48 at Lawrence Livermore National Laboratory, under
grant DE-FG02-97ER41029 at the University of Florida, and by an IBM Einstein 
Endowed
Fellowship at the Institute for Advanced Study.

\end{document}